\documentclass[12pt]{article}

\newcommand{\lsim}{
\mathrel{\hbox{\rlap{\hbox{\lower4pt\hbox{$\sim$}}}\hbox{$<$}}}}
\newcommand{\gsim}{
\mathrel{\hbox{\rlap{\hbox{\lower4pt\hbox{$\sim$}}}\hbox{$>$}}}}

\def\spose#1{\hbox to 0pt{#1\hss}}
\def\lsim{\mathrel{\spose{\lower 3pt\hbox{$\mathchar"218$}}
 \raise 2.0pt\hbox{$\mathchar"13C$}}}
\def\gsim{\mathrel{\spose{\lower 3pt\hbox{$\mathchar"218$}}
 \raise 2.0pt\hbox{$\mathchar"13E$}}}

\begin{document}

\begin{titlepage}

\begin{flushright}
CERN-TH/2003-010\\
hep-ph/0301255
\end{flushright}

\vspace{2cm}
\begin{center}
\boldmath
\large\bf New, Efficient and Clean Strategies to Explore\\ 
\vspace{0.3truecm}
CP Violation Through Neutral $B$ Decays
\unboldmath
\end{center}

\vspace{1.2cm}
\begin{center}
Robert Fleischer\\[0.1cm]
{\sl Theory Division, CERN, CH-1211 Geneva 23, Switzerland}
\end{center}

\vspace{1.7cm}
\begin{abstract}
\vspace{0.2cm}\noindent
We point out that decays of the kind $B_d\to D_\pm K_{\rm S (L)}$ and 
$B_s\to D_\pm \eta^{(')}, D_\pm \phi, ...$, where $D_+$ and $D_-$ denote the 
CP-even and CP-odd eigenstates of the neutral $D$-meson system, respectively, 
provide very efficient, theoretically clean determinations of the angle 
$\gamma$ of the unitarity triangle. In this new strategy, we use the 
$B^0_q$--$\overline{B^0_q}$ ($q\in\{d,s\}$) mixing phase $\phi_q$ as
an input, and employ only ``untagged'' and mixing-induced CP-violating 
observables, which satisfy a very simple relation, allowing us to determine 
$\tan\gamma$. Using a plausible dynamical assumption, $\gamma$ can be fixed 
in an essentially {\it unambiguous} manner. The corresponding formalism can
also be applied to $B_d\to D_\pm\pi^0,D_\pm\rho^0, ...$\ and 
$B_s\to D_\pm K_{\rm S (L)}$ decays. Although these modes appear less 
attractive for the extraction of $\gamma$, they provide interesting 
determinations of $\sin\phi_q$. In comparison with the conventional 
$B_d\to J/\psi K_{\rm S (L)}$ and $B_s\to J/\psi\phi$ methods, these 
extractions do not suffer from any penguin uncertainties, and are 
theoretically cleaner by one order of magnitude.
\end{abstract}

\vfill
\noindent
CERN-TH/2003-010\\
January 2003

\end{titlepage}

\thispagestyle{empty}
\vbox{}
\newpage
 
\setcounter{page}{1}

\section{Introduction}\label{sec:intro}
After the discovery of CP violation in the $B$-meson system by the
BaBar (SLAC) and Belle (KEK) collaborations \cite{CP-B-obs}, the 
exploration of CP violation is now entering another exciting stage,
where the central target is the unitarity triangle of the 
Cabibbo--Kobayashi--Maskawa (CKM) matrix, with its three angles 
$\alpha$, $\beta$ and $\gamma$. The goal is to overconstrain this
triangle as much as possible, thereby performing a stringent test 
of the Kobayashi--Maskawa \cite{KM} mechanism of CP violation (for 
a detailed review, see \cite{RF-PHYS-REP}).

Using the ``gold-plated'' mode $B_d\to J/\psi K_{\rm S}$ and similar 
channels, the efforts at the $B$ factories have already led to the
determination of $\sin\phi_d$ with an impressive accuracy, where 
$\phi_d$ is the CP-violating weak $B^0_d$--$\overline{B^0_d}$ mixing 
phase, which equals $2\beta$ in the Standard Model. The present world 
average is given by $\sin\phi_d=0.734\pm0.054$ \cite{nir}, implying
\begin{equation}
\phi_d=\left(47^{+5}_{-4}\right)^\circ \, \lor \, 
\left(133^{+4}_{-5}\right)^\circ.
\end{equation}
Here the former solution would be in perfect agreement with the ``indirect'' 
range implied by the CKM fits, $40^\circ\lsim\phi_d\lsim60^\circ$ 
\cite{CKM-fits}, whereas the latter would correspond to new physics. 
Measuring the sign of $\cos\phi_d$, both solutions can be distinguished.
There are several strategies on the market to accomplish this important task 
\cite{ambig}. In the $B\to J/\psi K$ system, $\mbox{sgn}(\cos\phi_d)$ can 
be extracted from the time-dependent angular distribution of the decay 
products of $B_d\to J/\psi[\to\ell^+\ell^-] K^\ast[\to\pi^0K_{\rm S}]$, 
if the sign of a hadronic parameter $\cos\delta_f$, involving a strong 
phase $\delta_f$, is fixed through factorization \cite{DDF2,DFN}. This 
analysis is already in progress at the $B$ factories \cite{itoh}. 

A key element in the testing of the Standard-Model description of 
CP violation is the determination of the angle $\gamma$ of the unitarity
triangle. In this context, $B\to\pi K,\pi\pi$ modes are receiving a lot 
of attention (for recent reviews, see \cite{BpiK-overviews}). The
theoretical accuracy of these approaches is mainly limited by 
non-factorizable $SU(3)$-breaking corrections. On the other hand, 
there are also certain classes of pure ``tree'' decays, allowing -- at
least in principle -- theoretically clean determinations of $\gamma$ (see, 
for instance, \cite{GroLo}--\cite{gro-BDK-02}). Unfortunately, the practical
implementation of these approaches is challenging, and the extraction
of $\gamma$ usually suffers from multiple discrete ambiguities, reducing
the power to search for possible signals of new physics significantly. 

The focus of the present paper is given by particularly interesting
colour-suppressed neutral $B_q$-meson decays ($q\in\{d,s\}$), which 
receive no contributions from penguin-like topologies. In 
Section~\ref{sec:b-s}, we shall first turn to decays of the kind 
$B_d^0\to D_\pm K_{\rm S (L)}$ and $B_s^0\to D_\pm \eta^{(')}, D_\pm \phi,...$,
which originate from $\overline{b}\to\overline{u}c\overline{s}$, 
$\overline{c} u \overline{s}$ quark-level processes, and point out that
they provide very efficient, theoretically clean determinations 
of $\gamma$ that are essentially free from discrete ambiguities (for 
alternative approaches using such modes to extract weak phases, see 
\cite{GroLo,KaLo,AtSo}). As usual, $D_+$ and $D_-$ are the CP-even and 
CP-odd eigenstates of the neutral $D$-meson system, respectively. As we 
will see in Section~\ref{sec:b-d}, the corresponding formalism can also 
be applied straightforwardly to decays of the kind 
$B_d^0\to D_\pm\pi^0, D_\pm\rho^0, ...$\ and $B_s^0\to D_\pm K_{\rm S (L)}$,
which arise from $\overline{b}\to \overline{u} c \overline{d}$, 
$\overline{c} u \overline{d}$ quark-level transitions. Since the extraction 
of $\gamma$ relies on certain interference effects, which are doubly 
Cabibbo-suppressed in these modes, they appear not as attractive as the 
$B_d^0\to D_\pm K_{\rm S (L)}$ and $B_s^0\to D_\pm \eta^{(')}, D_\pm \phi, 
...$\ channels. However, they provide interesting extractions of 
$\sin\phi_q$, where $\phi_q$ is the $B^0_q$--$\overline{B^0_q}$ mixing phase. 
In comparison with the conventional $B_d\to J/\psi K_{\rm S (L)}$ and 
$B_s\to J/\psi\phi$ methods to determine these quantities, these new
strategies do not suffer from any penguin uncertainties, and are theoretically 
cleaner by one order of magnitude. Finally, we summarize our conclusions in
Section~\ref{sec:concl}.

\boldmath
\section{$B_d\to D_\pm K_{\rm S (L)}$ and 
$B_s\to D_\pm \eta^{(')}, D_\pm \phi, ...$}\label{sec:b-s}
\unboldmath
In our analysis, we shall neglect $D^0$--$\overline{D^0}$ mixing and 
CP violation in $D$ decays, which are tiny effects in the Standard 
Model; should they be enhanced through new physics, they could be taken
into account with the help of experimental $D$-decay studies \cite{D-studies}.
We may then write
\begin{equation}\label{D-CP-def}
|D_\pm\rangle=\frac{1}{\sqrt{2}}\left[|D^0\rangle\pm e^{i\phi_{\rm CP}(D)}
|\overline{D^0}\rangle\right],
\end{equation}
where 
\begin{equation}\label{phi-D-def}
({\cal CP})|D^0\rangle=e^{i\phi_{\rm CP}(D)}|\overline{D^0}\rangle,
\quad
({\cal CP})|\overline{D^0}\rangle=e^{-i\phi_{\rm CP}(D)}|D^0\rangle.
\end{equation}
In order to simplify the following discussion, let us denote the 
$B_d^0\to D_\pm K_{\rm S (L)}$ and 
$B_s^0\to D_\pm \eta^{(')}, D_\pm \phi,...$\ decays generically by 
$B_q^0\to D_\pm f_s$, where the label $s$ reminds us that we are 
dealing with $\overline{b}\to\overline{s}$ transitions, and $f_s$ is
a CP eigenstate, satisfying
\begin{equation}\label{CP-fs}
({\cal CP})|f_s\rangle=\eta_{\rm CP}^{f_s}|f_s\rangle.
\end{equation}

\boldmath
\subsection{Untagged Observables and a New Bound on 
$\gamma$}\label{subsec:untagged}
\unboldmath
The most straightforward observable we may consider is the 
``untagged'' rate 
\begin{equation}
\langle\Gamma(B_q(t)\to D_\pm f_s)\rangle\equiv
\Gamma(B^0_q(t)\to D_\pm f_s)+\Gamma(\overline{B^0_q}(t)\to D_\pm f_s),
\end{equation}
where $\Gamma(B^0_q(t)\to D_\pm f_s)$ and $\Gamma(\overline{B^0_q}(t)\to 
D_\pm f_s)$ denote the time-dependent decay rates for initially, i.e.\ at 
time $t=0$, present $B^0_q$ and $\overline{B^0_q}$ states, respectively.
It takes the following form \cite{RF-PHYS-REP}:
\begin{eqnarray}
\lefteqn{\langle\Gamma(B_q(t)\to D_\pm f_s)\rangle=
\left[\Gamma(B^0_q\to D_\pm f_s)+\Gamma(\overline{B^0_q}\to D_\pm f_s)
\right]}\nonumber\\
&&\times\left[\cosh(\Delta\Gamma_qt/2)-{\cal A}_{\rm \Delta\Gamma}
(B_q\to D_\pm f_s)\,\sinh(\Delta\Gamma_qt/2)\right]
e^{-\Gamma_qt},\label{untagged}
\end{eqnarray}
where $\Delta\Gamma_q\equiv\Gamma_{\rm H}^{(q)}-\Gamma_{\rm L}^{(q)}$ is
the decay width difference of the $B_q$ mass eigenstates $B_q^{\rm H}$ 
(``heavy'') and $B_q^{\rm L}$ (``light''), and 
$\Gamma_q\equiv(\Gamma_{\rm H}^{(q)}+\Gamma_{\rm L}^{(q)})/2$ is their 
average decay width. In the case of the $B_d$-meson system, the width
difference is negligibly small, so that the time evolution of (\ref{untagged}) 
is essentially given by the well-known exponential $e^{-\Gamma_dt}$. On the 
other hand, $\Delta \Gamma_s$ may be as large as ${\cal O}(-10\%)$ (for a 
recent review, see \cite{BeLe}). Strategies employing $\Delta\Gamma_s$ 
to extract $\gamma$ were proposed in \cite{DF-untagged}. Since we do not 
have to rely on a sizeable value of $\Delta\Gamma_q$ and the observable 
${\cal A}_{\rm \Delta\Gamma}(B_q\to D_\pm f_s)$ in the following analysis, 
we shall not consider these quantities in further detail; the effects of 
$\Delta\Gamma_q$ could be taken into account straightforwardly. 

In the new strategy to determine $\gamma$ discussed below, we apply
(\ref{untagged}) only to extract the ``unevolved'', i.e.\ time-independent, 
untagged rates 
\begin{equation}
\langle\Gamma(B_q\to D_\pm f_s)\rangle\equiv\Gamma(B^0_q\to D_\pm f_s)+
\Gamma(\overline{B^0_q}\to D_\pm f_s),
\end{equation}
which allow us to determine the following asymmetry:
\begin{equation}\label{Gam-pm-def}
\Gamma_{+-}^{f_s}\equiv
\frac{\langle\Gamma(B_q\to D_+ f_s)\rangle-\langle
\Gamma(B_q\to D_- f_s)\rangle}{\langle\Gamma(B_q\to D_+ f_s)\rangle
+\langle\Gamma(B_q\to D_- f_s)\rangle}.
\end{equation}
This quantity will play a key r\^ole below. Using (\ref{D-CP-def}), we 
obtain
\begin{equation}
A(B^0_q\to D_\pm f_s)=\frac{1}{\sqrt{2}}\left[A(B^0_q\to D^0f_s)
\pm e^{-i\phi_{\rm CP}(D)}A(B^0_q\to \overline{D^0}f_s)\right]
\end{equation}
\begin{equation}
A(\overline{B^0_q}\to D_\pm f_s)=\frac{1}{\sqrt{2}}\left[
A(\overline{B^0_q}\to D^0f_s)\pm e^{-i\phi_{\rm CP}(D)}
A(\overline{B^0_q}\to \overline{D^0}f_s)\right].
\end{equation}
If we follow \cite{RF-PHYS-REP}, and employ an appropriate low-energy 
effective Hamiltonian to deal with the $A(B_q\to D f_s)$ decay amplitudes,
we eventually arrive at 
\begin{equation}\label{Gam-pm}
\Gamma_{+-}^{f_s}=\frac{2\,x_{f_s}\cos\delta_{f_s}\cos\gamma}{1+x_{f_s}^2}.
\end{equation}
Here $\gamma$ is the usual angle of the unitarity triangle, and
\begin{equation}\label{x-def}
x_{f_s}e^{i\delta_{f_s}}\equiv R_b e^{-i\phi_{\rm CP}(D)}
\left[\frac{\langle f_s\overline{D^0}|
{\cal O}_1^s C_1(\mu)+{\cal O}_2^s C_2(\mu)|
\overline{B^0_q}\rangle}{\langle f_sD^0|
\overline{{\cal O}}_1^s C_1(\mu)+
\overline{{\cal O}}_2^s C_2(\mu)|
\overline{B^0_q}\rangle}\right]
\end{equation}
denotes a hadronic parameter, which involves the current--current operators
\begin{equation}\label{op-basis}
\begin{array}{rclrcl}
{\cal O}_1^s&=&(\overline{s}_\alpha c_\beta)_{\mbox{{\scriptsize V--A}}}
\left(\overline{u}_\beta b_\alpha\right)_{\mbox{{\scriptsize V--A}}},&
~~{\cal O}_2^s
&=&(\overline{s}_\alpha c_\alpha)_{\mbox{{\scriptsize V--A}}}
\left(\overline{u}_\beta b_\beta\right)_{\mbox{{\scriptsize V--A}}},\\
&&&&&\\
\overline{{\cal O}}_1^s
&=&(\overline{s}_\alpha u_\beta)_{\mbox{{\scriptsize 
V--A}}}\left(\overline{c}_\beta b_\alpha\right)_{\mbox{{\scriptsize V--A}}},&
~~\overline{{\cal O}}_2^s
&=&(\overline{s}_\alpha u_\alpha)_{\mbox{{\scriptsize 
V--A}}}\left(\overline{c}_\beta b_\beta\right)_{\mbox{{\scriptsize V--A}}},\\
\end{array}
\end{equation}
with their Wilson coefficients $C_{1,2}(\mu)$, and the 
CKM factor \cite{Andr-02}
\begin{equation}
R_b\equiv\left(1-\frac{\lambda^2}{2}\right)\frac{1}{\lambda}\left|
\frac{V_{ub}}{V_{cb}}\right|=0.39\pm0.04,
\end{equation}
where $\lambda\equiv|V_{us}|=0.22$ is the usual Wolfenstein parameter 
\cite{wolf}. 

If we apply the factorization approach to calculate (\ref{x-def})
and perform appropriate CP transformations by taking (\ref{phi-D-def}) 
into account, we observe that both the convention-dependent phase 
$\phi_{\rm CP}(D)$ and the factorized hadronic matrix elements cancel, 
and arrive at
\begin{equation}\label{x-fact}
\left.x_{f_s}e^{i\delta_{f_s}}\right|_{\rm fact}=-R_b\approx -0.4.
\end{equation}
Note that $x_{f_s}e^{i\delta_{f_s}}$ is governed by a {\it ratio} of hadronic 
matrix elements, which are related to each other through an interchange of 
all up and charm quarks, hence having a similar structure, and that 
$\delta_{f_s}$ measures the {\it relative} strong phase between them. 
Consequently, we expect that the deviation of $\delta_{f_s}$ from the
trivial value of $180^\circ$ is moderate, even if the individual hadronic 
matrix elements entering (\ref{x-def}) should deviate sizeably from the 
factorization case, as advocated in \cite{non-fact}. In particular, the 
assumption 
\begin{equation}\label{dyn-assumpt}
\cos\delta_{f_s}<0,
\end{equation}
which is satisfied for the whole range of $90^\circ<\delta_{f_s}<270^\circ$, 
appears very plausible.

An important advantage of the observable $\Gamma_{+-}^{f_s}$ is that it does 
not depend on the overall normalization of the untagged 
$\langle\Gamma(B_q\to D_\pm f_s)\rangle$ rates. Interestingly, already
$\Gamma_{+-}^{f_s}$ provides valuable information on $\gamma$. Using
(\ref{Gam-pm}), we obtain
\begin{equation}\label{gam-bound}
|\cos\gamma|\geq|\Gamma_{+-}^{f_s}|,
\end{equation}
which can easily be converted into bounds on $\gamma$ (for alternative
constraints on $\gamma$ that could be obtained from other pure tree 
decays involving $D_\pm$ states, see \cite{gro-BDK,gro-BDK-02}). Note 
that $\Gamma_{+-}^{f_s}$ satisfies, by definition, the relation
$0\leq|\Gamma_{+-}^{f_s}|\leq1$. Moreover, we have
\begin{equation}
\mbox{sgn}(\Gamma_{+-}^{f_s})=\mbox{sgn}(\cos\delta_{f_s}) \,
\mbox{sgn}(\cos\gamma).
\end{equation}
Using now (\ref{dyn-assumpt}), i.e.\ $\mbox{sgn}(\cos\delta_{f_s})=-1$, 
we may fix the sign of $\cos\gamma$ through the sign of
$\Gamma_{+-}^{f_s}$ as follows:
\begin{equation}\label{sgn-gam}
\mbox{sgn}(\cos\gamma)=-\mbox{sgn}(\Gamma_{+-}^{f_s}).
\end{equation}
We shall come back to this interesting feature below.

\boldmath
\subsection{Tagged Observables and Extraction of 
$\gamma$}\label{subsec:tagged}
\unboldmath
The final goal is not just to constrain $\gamma$, but to
{\it determine} this angle. To this end, we use the following 
time-dependent rate asymmetry \cite{RF-PHYS-REP}:
\begin{eqnarray}
\lefteqn{\frac{\Gamma(B^0_q(t)\to D_\pm f_s)-
\Gamma(\overline{B^0_q}(t)\to D_\pm f_s)}{\Gamma(B^0_q(t)\to D_\pm f_s)+
\Gamma(\overline{B^0_q}(t)\to D_\pm f_s)}}\label{ee6}\\
&&=\left[\frac{{\cal A}_{\rm CP}^{\rm dir}(B_q\to D_\pm f_s)\cos(\Delta M_q t)+
{\cal A}_{\rm CP}^{\rm mix}(B_q\to D_\pm f_s)
\sin(\Delta M_q t)}{\cosh(\Delta\Gamma_qt/2)-{\cal A}_{\rm 
\Delta\Gamma}(B_q\to D_\pm f_s)\sinh(\Delta\Gamma_qt/2)}\right],\nonumber
\end{eqnarray}
where $\Delta M_q\equiv M_{\rm H}^{(q)}-M_{\rm L}^{(q)}>0$ denotes the 
mass difference between the $B_q$ mass eigenstates. Let us note that 
${\cal A}_{\rm \Delta\Gamma}(B_q\to D_\pm f_s)$, which could be 
extracted from the untagged rate (\ref{untagged}) in the presence of a 
sizeable $\Delta\Gamma_q$, is not independent from 
${\cal A}_{\rm CP}^{\rm dir}(B_q\to D_\pm f_s)$ and 
${\cal A}_{\rm CP}^{\rm mix}(B_q\to D_\pm f_s)$, satisfying
\begin{equation}\label{Obs-rel}
\Bigl[{\cal A}_{\rm CP}^{\rm dir}(B_q\to D_\pm f_s)\Bigr]^2+
\Bigl[{\cal A}_{\rm CP}^{\rm mix}(B_q\to D_\pm f_s)\Bigr]^2+
\Bigl[{\cal A}_{\Delta\Gamma}(B_q\to D_\pm f_s)\Bigr]^2=1.
\end{equation}

As discussed in \cite{RF-PHYS-REP}, the ``direct'' CP-violating observables 
${\cal A}_{\rm CP}^{\rm dir}$ originate from interference effects between 
different decay amplitudes. In the case of $\overline{B_q^0}\to D_\pm f_s$
transitions, these are provided by the $\overline{B_q^0}\to 
\overline{D^0} f_s$ and $\overline{B_q^0}\to D^0 f_s$ decay paths, yielding
\begin{equation}
C_\pm^{f_s}\equiv {\cal A}_{\rm CP}^{\rm dir}(B_q\to D_\pm f_s)=
\mp\left[\frac{2\,x_{f_s}\sin\delta_{f_s}\sin\gamma}{1\pm
2\,x_{f_s}\cos\delta_{f_s}\cos\gamma+x_{f_s}^2}\right].
\end{equation}
It is convenient to consider the following combinations:
\begin{equation}\label{Cp-def}
\langle C_{f_s}\rangle_+\equiv\frac{C_+^{f_s}+C_-^{f_s}}{2}
=\frac{x_{f_s}^2\sin2\delta_{f_s}\sin2\gamma}{(1+x_{f_s}^2)^2-(2\,x_{f_s}
\cos\delta_{f_s}\cos\gamma)^2}
\end{equation}
\begin{equation}\label{Cm-def}
\langle C_{f_s}\rangle_-\equiv\frac{C_+^{f_s}-C_-^{f_s}}{2}
=-\left[\frac{2\,x_{f_s}(1+x_{f_s}^2)\sin\delta_{f_s}
\sin\gamma}{(1+x_{f_s}^2)^2-(2\,x_{f_s}\cos\delta_{f_s}\cos\gamma)^2}\right],
\end{equation}
which give 
\begin{equation}
-\left[\frac{\langle C_{f_s}\rangle_+}{\langle C_{f_s}\rangle_-}\right]=
\frac{2\,x_{f_s}\cos\delta_{f_s}\cos\gamma}{1+x_{f_s}^2}=\Gamma_{+-}^{f_s}.
\end{equation}
Consequently, (\ref{Cp-def}) and (\ref{Cm-def}) are not independent from 
$\Gamma_{+-}^{f_s}$, and allow us to extract the same information. 
However, the ``untagged'' avenue offered by (\ref{Gam-pm-def}) is much more 
promising for the determination of $\Gamma_{+-}^{f_s}$ from a practical 
point of view. Note that the $C_\pm^{f_s}$ vanish in the case of 
$\delta_{f_s}=180^\circ$, since these observables are proportional to 
$\sin\delta_{f_s}$.  

The phase $\delta_{f_s}$ can be probed nicely through the relation 
\begin{equation}\label{tan-delta}
\tan\delta_{f_s} \tan\gamma =-\left[\frac{1-
(\Gamma_{+-}^{f_s})^2}{\Gamma_{+-}^{f_s}}\right]\langle C_{f_s}\rangle_-,
\end{equation}
which follows from the elimination of $x_{f_s}$ in (\ref{Cm-def}) 
through (\ref{Gam-pm}). As we will see below, $\tan\gamma$ can be determined
straightforwardly, thereby allowing the extraction of $\tan\delta_{f_s}$,
which implies a twofold solution for $\delta_{f_s}$ itself. If we use the
plausible assumption (\ref{dyn-assumpt}), we may resolve this ambiguity,
and arrive at a single solution for $\delta_{f_s}$, which offers valuable 
insights into hadronic physics. 

Let us now turn to the mixing-induced CP-violating observables, which 
are associated with the $\sin(\Delta M_q t)$ terms in (\ref{ee6}), and 
originate from interference effects between $B^0_q$--$\overline{B^0_q}$ 
mixing and decay processes. Following the formalism discussed in 
\cite{RF-PHYS-REP}, we obtain
\begin{equation}\label{Spm-expr}
S_\pm^{f_s}\equiv {\cal A}_{\rm CP}^{\rm mix}(B_q\to D_\pm f_s)=\pm\,\eta_{f_s}
\left[\frac{\sin\phi_q \pm 2\,x_{f_s}\cos\delta_{f_s}\sin(\phi_q+\gamma)+
x_{f_s}^2\sin(\phi_q+2\gamma)}{1 \pm 2\,x_{f_s}\cos\delta_{f_s}\cos\gamma
+x_{f_s}^2}\right],
\end{equation}
where the factor
\begin{equation}
\eta_{f_s}\equiv(-1)^L\eta_{\rm CP}^{f_s}
\end{equation}
takes into account both the angular momentum $L$ of the $Df_s$ state 
and the intrinsic CP parity $\eta_{\rm CP}^{f_s}$ of $f_s$ (see 
(\ref{CP-fs})), and
\begin{equation}\label{phi-q-def}
\phi_q \equiv 2\,\mbox{arg}(V_{tq}^\ast V_{tb})\stackrel{\rm SM}{=}
\left\{\begin{array}{cl}
+2\beta={\cal O}(50^\circ) & ~~\mbox{($q=d$)}\\
-2\lambda^2\eta={\cal O}(-2^\circ) & ~~\mbox{($q=s$)}
\end{array}\right.
\end{equation}
is the CP-violating weak $B^0_q$--$\overline{B^0_q}$ mixing phase; the
quantity $\eta$ appearing in the $B_s$ case is another Wolfenstein 
parameter \cite{wolf}. In analogy to (\ref{Cp-def}) and (\ref{Cm-def}), 
it is convenient to consider the following combinations:
\begin{equation}\label{Sp-def}
\langle S_{f_s}\rangle_+\equiv\frac{S_+^{f_s}+S_-^{f_s}}{2}
=\eta_{f_s}\left[\frac{2\,x_{f_s}\cos\delta_{f_s}\sin\gamma
\left\{\cos\phi_q-x_{f_s}^2\cos(\phi_q+2\gamma)\right\}}{(1+x_{f_s}^2)^2-
(2\,x_{f_s}\cos\delta_{f_s}\cos\gamma)^2}\right]
\end{equation}
\begin{eqnarray}
\lefteqn{\langle S_{f_s}\rangle_-\equiv\frac{S_+^{f_s}-S_-^{f_s}}{2}}
\label{Sm-def}\\
&&=\eta_{f_s}\left[\frac{\sin\phi_q+x_{f_s}^2\left\{\sin\phi_q+
(1+x_{f_s}^2)\sin(\phi_q+2\gamma)-4\cos^2\delta_{f_s}
\cos\gamma\sin(\phi_q+\gamma)\right\}}{(1+x_{f_s}^2)^2-
(2\,x_{f_s}\cos\delta_{f_s}\cos\gamma)^2}\right].\nonumber
\end{eqnarray}
An important advantage in comparison with $\langle C_{f_s}\rangle_+$ and 
$\langle C_{f_s}\rangle_-$ is that $\delta_{f_s}$ now enters only in the form 
of $\cos\delta_{f_s}$. Although (\ref{Sp-def}) and (\ref{Sm-def}) are 
complicated expressions, we may use (\ref{Gam-pm}) to derive the following, 
very simple final result:
\begin{equation}\label{final-result}
\tan\gamma\cos\phi_q=
\left[\frac{\eta_{f_s} \langle S_{f_s}\rangle_+}{\Gamma_{+-}^{f_s}}\right]
+\left[\eta_{f_s}\langle S_{f_s}\rangle_--\sin\phi_q\right].
\end{equation}
It should be emphasized that this relation is valid {\it exactly}. In 
particular, it does not rely on any assumptions related to factorization or 
the strong phase $\delta_{f_s}$. Interestingly, the first term in square 
brackets can be considered as the leading ${\cal O}(1)$ term, as it is 
determined from a ratio of observables that are both governed by 
$x_{f_s}\approx0.4$. The second term starts to contribute at 
${\cal O}(x_{f_s}^2)$, as can easily be seen from (\ref{Sm-def}).

\boldmath
\subsection{Resolving Discrete Ambiguities}\label{subsec:ambig-res}
\unboldmath
The simple and completely general relation given in (\ref{final-result}) 
has powerful applications. If we assume that $\phi_q$ will be known 
unambiguously with the help of \cite{ambig}--\cite{DFN} by the time the 
$B_q\to D_\pm f_s$ measurements can be performed in practice, 
(\ref{final-result}) allows us to determine $\tan\gamma$ unambiguously, 
yielding a twofold solution, $\gamma=\gamma_1\,\lor\gamma_2$, where we 
may choose $\gamma_1\in[0^\circ,180^\circ]$ and $\gamma_2=\gamma_1+180^\circ$. 
If we assume -- as is usually done -- that $\gamma$ lies between $0^\circ$ 
and $180^\circ$, we may exclude the $\gamma_2$ solution. The range 
$[0^\circ,180^\circ]$ for $\gamma$ is implied by the Standard-Model 
interpretation of $\varepsilon_K$, which measures the ``indirect'' 
CP violation in the neutral kaon system, if we make the very 
plausible assumption that a certain ``bag'' parameter, $B_K$, is positive. 
Let us note that we have also assumed implicitly in (\ref{phi-q-def}) 
that another ``bag'' parameter $B_{B_q}$, which is the $B_q$-meson counterpart 
of $B_K$, is positive as well. Indeed, all existing non-perturbative 
methods give positive values for these parameters. For a discussion of 
the very unlikely $B_K<0$, $B_{B_q}<0$ cases, see \cite{GKN}. 

If we assume that only $\sin\phi_q$ has been measured through mixing-induced 
CP-violating effects, we may use the right-hand side of (\ref{final-result}) 
to determine the sign of the product $\tan\gamma \cos\phi_q$. As we have 
seen above, $\sin\phi_d$ has already been determined with an impressive 
experimental accuracy due to the efforts at the $B$ factories. 
Since $\mbox{sgn}(\cos\gamma)=\mbox{sgn}(\tan\gamma)$ 
for $\gamma\in[0^\circ,180^\circ]$, we may use (\ref{sgn-gam}) to
fix the sign of $\tan\gamma$, and may eventually determine the sign of 
$\cos\phi_q=\pm\sqrt{1-\sin^2\phi_q}$, thereby resolving
the twofold ambiguity arising in the extraction of $\phi_q$ from
$\sin\phi_q$. Consequently, measuring the untagged asymmetry 
$\Gamma_{+-}^{f_s}$ and the mixing-induced observables 
$\langle S_{f_s}\rangle_+$ and $\langle S_{f_s}\rangle_-$ of 
$B_d\to D_\pm K_{\rm S (L)}$ modes, we may fix $\phi_d$ and 
$\gamma\in[0^\circ,180^\circ]$ in an unambiguous way. Following these 
lines, we may also distinguish between the two cases of 
$(\phi_d,\gamma)\sim(47^\circ,60^\circ)$ and $(133^\circ,120^\circ)$,
which emerge from a recent analysis of CP violation in $B_d\to\pi^+\pi^-$
\cite{FlMa2}. 

Since the expectation $\gamma\in[0^\circ,180^\circ]$ relies on the
Standard-Model interpretation of $\varepsilon_K$, it may no longer be
correct in the presence of new physics. Consequently, it would be
very interesting to check whether $\gamma$ is actually smaller than 
$180^\circ$. If we assume that $\phi_q$ is known unambiguously, 
(\ref{final-result}) implies the twofold solution 
$\gamma=\gamma_1\,\lor\gamma_2$, where $\gamma_1\in[0^\circ,180^\circ]$ 
and $\gamma_2=\gamma_1+180^\circ$, as we have seen above. Since 
$\cos\gamma_1$ and $\cos\gamma_2$ have {\it opposite} signs, we may easily
distinguish between these solutions with the help of (\ref{sgn-gam}).

\boldmath
\subsection{Special Cases Related to $\Gamma_{+-}^{f_s}=0$}\label{subsec:spec}
\unboldmath
Let us, for completeness, also spend a few words on two special cases, 
where $\Gamma_{+-}^{f_s}$ vanishes. The first one corresponds to 
$\gamma=90^\circ \lor 270^\circ$, yielding
\begin{equation}\label{Sp-special}
\left.\eta_{f_s}\langle S_{f_s}\rangle_+\right|_{\gamma=90^\circ 
\lor 270^\circ}
=\left[\frac{2\,x_{f_s}\cos\delta_{f_s}\cos\phi_q}{1+x_{f_s}^2}\right]
\mbox{sgn}(\sin\gamma).
\end{equation}
Consequently, if we employ again (\ref{dyn-assumpt}) and assume that the 
sign of $\cos\phi_q$ will be known by the time the $B_q\to D_\pm f_s$ 
measurements can be performed in practice, this observable allows us to 
determine the sign of $\sin\gamma$, thereby distinguishing between 
$\gamma=90^\circ$ and $270^\circ$. Concerning the extraction of 
$\delta_{f_s}$ (see (\ref{tan-delta})), we may now use the ratio of
\begin{equation}
\left.\langle C_{f_s}\rangle_-\right|_{\gamma=90^\circ \lor 270^\circ}
=-\left[\frac{2\,x_{f_s}\sin\delta_{f_s}}{1+x_{f_s}^2}\right]
\mbox{sgn}(\sin\gamma)
\end{equation}
and (\ref{Sp-special}) to determine $-\tan\delta_{f_s}/\cos\phi_q$. 

In the unlikely case of $\delta_{f_s}=90^\circ \lor 270^\circ$, which is 
the second possibility yielding $\Gamma_{+-}^{f_s}=0$, 
$\langle S_{f_s}\rangle_+$ would vanish as well. However, we could then 
employ
\begin{equation}
\left.\eta_{f_s}\langle S_{f_s}\rangle_-\right|_{\delta_{f_s}=90^\circ 
\lor 270^\circ}
=\sin\phi_q+\frac{2\,x_{f_s}^2}{1+x_{f_s}^2}\sin\gamma\cos(\phi_q+\gamma)
\end{equation}
and
\begin{equation}
\left.\langle C_{f_s}\rangle_-\right|_{\delta_{f_s}=90^\circ 
\lor 270^\circ}=-\left[\frac{2\,x_{f_s}\sin\gamma}{1+x_{f_s}^2}\right]
\mbox{sgn}(\sin\delta_{f_s})
\end{equation}
to determine $\gamma$ and $x_{f_s}$. Unfortunately, we would then have
to struggle with discrete ambiguities, and would have to rely on the
experimental resolution of terms entering at the $x_{f_s}^2$ level. 
Note that the value of $x_{f_s}$ could differ dramatically from 
$R_b\approx0.4$ in the case of $\delta_{f_s}=90^\circ \lor 270^\circ$.

\boldmath
\subsection{Remarks on 
$B_s\to D_\pm \eta^{(')}, D_\pm \phi, ...$}\label{subsec:Bs}
\unboldmath
Considering decays of the kind $B_s\to D_\pm \eta^{(')}, D_\pm \phi, ...$,
we may implement the strategy discussed above in the $B_s$-meson system. 
These transitions are not accessible at the asymmetric $e^+e^-$ $B$ factories 
operating at the $\Upsilon(4S)$ resonance, i.e.\ they cannot be measured by 
the BaBar and Belle collaborations, but may be studied at hadronic $B$ 
experiments, in particular at LHCb (CERN) and BTeV (Fermilab). Since $\phi_s$ 
is negligibly small in the Standard Model, yielding $\sin\phi_s=0$ and 
$\cos\phi_s=+1$, (\ref{final-result}) simplifies even further in this 
interesting case:
\begin{equation}\label{final-result-Bs}
\left.\tan\gamma\right|_{B_s}^{\rm SM}=
\frac{\eta_{f_s} \langle S_{f_s}\rangle_+}{\Gamma_{+-}^{f_s}}
+\eta_{f_s}\langle S_{f_s}\rangle_-.
\end{equation}

As is well known, $\phi_s$ is a sensitive probe for new-physics 
contributions to $B^0_s$--$\overline{B^0_s}$ mixing, which may lead 
to a sizeable value of $\phi_s$. In addition to discrepancies between 
the values for $\gamma$ extracted from (\ref{final-result-Bs}) on the one 
hand, and its $B_d$ counterpart on the other hand, such scenarios may also 
be probed through the search for CP violation in 
$B_s\to J/\psi \phi$ modes. If we use these channels (or those
proposed below) to fix $\phi_s$, we may extract the ``true'' value of 
$\gamma$, and may resolve the discrete ambiguities as discussed above. 

In this game, it is also useful to combine information provided
by the $B_d$ and $B_s$ systems. For example, an interesting case arises
if actually no CP violation should be found in $B_s\to J/\psi \phi$, thereby
indicating $\sin\phi_s=0$. Then the question arises whether $\cos\phi_s$ is 
positive, as in the Standard Model, or negative, as in the case of
new physics. If we determine $\tan\gamma$ unambiguously through
$B_d\to D_\pm K_{\rm S (L)}$ modes as discussed above, we may use, 
for instance, $B_s\to D_\pm\phi$ to fix $\cos\phi_s$ with the help of 
(\ref{final-result}); for the extraction of the sign of $\cos\phi_s$,
it is of course sufficient to use only information on the sign of 
$\tan\gamma$ and the right-hand side of (\ref{final-result}) (alternative 
approaches to determine $\phi_s$ unambiguously can be found in \cite{DFN}).

\boldmath
\subsection{Numerical Examples}\label{subsec:num}
\unboldmath
In order to illustrate the strategies proposed in this paper in further 
detail, we consider $x_{f_s}=0.4$ and $\delta_{f_s}=180^\circ$, which 
corresponds to (\ref{x-fact}), and give in Table~\ref{tab:illu} the results 
for $\Gamma_{+-}^{f_s}$ and the relevant mixing-induced observables for 
various values of $\phi_q$ and $\gamma$. Since the strong phase 
$\delta_{f_s}$ enters in these observables only through $\cos\delta_{f_s}$, 
they are rather robust with respect to deviations of $\delta_{f_s}$ from
$180^\circ$. Playing with the numerical examples given in Table~\ref{tab:illu},
it is an easy exercise to see how the strategy proposed above and the 
resolution of the discrete ambiguities work in practice. Moreover,
it is interesting to observe that the first term in square brackets on the
right-hand side of (\ref{final-result}) is actually the leading one, and 
that the second term, which arises at the $x_{f_s}^2$ level, plays only a 
minor r\^ole. Let us also note that (\ref{gam-bound}) implies 
$-70^\circ\leq\gamma\leq70^\circ$ and $110^\circ\leq\gamma\leq250^\circ$ 
for $\Gamma_{+-}^{f_s}=-0.345$ and $+0.345$, respectively, where we
have also taken (\ref{sgn-gam}) into account.

\begin{table}
\begin{center}
\begin{tabular}{|c||c|c|c|c|c|c|c|}
\hline
$(\phi_q,\gamma)$ & $\Gamma_{+-}^{f_s}$ & $\eta_{f_s} S_+^{f_s}$ & 
$\eta_{f_s} S_-^{f_s}$ & $\eta_{f_s} \langle S_{f_s}\rangle_+$ &
$\eta_{f_s} \langle S_{f_s}\rangle_-$ & 1st term & 2nd term\\
\hline
\hline
$(47^\circ,60^\circ)$ & $-0.345$ & $+0.003$ & $-0.982$ & $-0.490$ &
$+0.493$ & $+1.420$ & $-0.239$\\ 
$(133^\circ,120^\circ)$ & $+0.345$ & $+0.982$ & $-0.003$ & $+0.490$ &
$+0.493$ & $+1.420$ & $-0.239$\\
$(47^\circ,240^\circ)$ & $+0.345$ & $+0.982$ & $-0.003$  & $+0.490$ &
$+0.493$ & $+1.420$ & $-0.239$\\ 
\hline
$(0^\circ,60^\circ)$ & $-0.345$ & $-0.729$ & $-0.533$ & $-0.631$ & 
$-0.098$ & $+1.830$ & $-0.098$\\ 
$(180^\circ,60^\circ)$ & $-0.345$ & $+0.729$ & $+0.533$ & $+0.631$ & 
$+0.098$ & $-1.830$ & $+0.098$\\
$(0^\circ,240^\circ)$ & $+0.345$ & $+0.533$ & $+0.729$ & $+0.631$ & 
$-0.098$ & $+1.830$ & $-0.098$\\ 
\hline
\end{tabular}
\caption{The relevant observables in the case of $x_{f_s}=0.4$ and 
$\delta_{f_s}=180^\circ$ for various values of $(\phi_q,\gamma)$. The 
columns ``1st term'' and ``2nd term'' refer to the 1st and 2nd terms 
in square brackets on the right-hand side of (\ref{final-result}), 
as discussed in the text.}\label{tab:illu}
\end{center}
\end{table}

\boldmath
\section{$B_d\to D_\pm\pi^0, D_\pm\rho^0, ...$\ and 
$B_s\to D_\pm K_{\rm S (L)}$}\label{sec:b-d}
\unboldmath
Making straightforward replacements of variables, the formalism 
developed above can also be applied to decays of the kind 
$B_d^0\to D_\pm\pi^0, D_\pm\rho^0, ...$\ and $B_s^0\to D_\pm K_{\rm S (L)}$,
which originate from $\overline{b}\to \overline{u} c \overline{d}$,
$\overline{c} u \overline{d}$ quark-level transitions. To this end, we
have only to substitute $x_{f_s}e^{i\delta_{f_s}}$ through
\begin{equation}\label{xd-def}
x_{f_d}e^{i\delta_{f_d}}=-\left(\frac{\lambda^2R_b}{1-\lambda^2}\right) 
e^{-i\phi_{\rm CP}(D)}\left[\frac{\langle f_d\overline{D^0}|
{\cal O}_1^d C_1(\mu)+{\cal O}_2^d C_2(\mu)|\overline{B^0_q}
\rangle}{\langle f_dD^0|\overline{{\cal O}}_1^d C_1(\mu)+
\overline{{\cal O}}_2^d C_2(\mu)|
\overline{B^0_q}\rangle}\right],
\end{equation}
yielding
\begin{equation}\label{xd-fact}
\left.x_{f_d}e^{i\delta_{f_d}}\right|_{\rm fact}=
+\left(\frac{\lambda^2R_b}{1-\lambda^2}\right)\approx 0.02.
\end{equation}
Note that the tiny value of $x_{f_d}$ in (\ref{xd-fact}) reflects that the 
interference effects between the $\overline{B^0_q}\to \overline{D^0} f_d$ and 
$\overline{B^0_q}\to D^0 f_d$ decay paths are doubly Cabibbo-suppressed, in 
contrast to the $B_q\to D f_s$ case. Consequently, since $\Gamma_{+-}^{f_d}$
and $\langle S_{f_d}\rangle_+$ are proportional to $x_{f_d}$, these 
observables are now strongly suppressed, taking values of at most a few 
per cent, and the strategy to determine $\gamma$ proposed above does not 
appear to be very attractive in this case. However, $\langle S_{f_d}\rangle_-$ 
provides an interesting extraction of $\sin\phi_q$ through
\begin{equation}\label{sin-phi-det}
\eta_{f_d}\langle S_{f_d}\rangle_-=\sin\phi_q + {\cal O}(x_{f_d}^2)
=\sin\phi_q + {\cal O}(4\times 10^{-4}).
\end{equation}
Note that the individual mixing-induced CP asymmetries $S_+^{f_d}$ and 
$S_-^{f_d}$ are affected by terms of ${\cal O}(x_{f_d})=
{\cal O}(2\times 10^{-2})$, as can be seen in (\ref{Spm-expr}). In the 
conventional strategies to determine $\sin\phi_d$ and $\sin\phi_s$ through 
$B_d\to J/\psi K_{\rm S (L)}$ and $B_s\to J/\psi \phi$, respectively, 
unknown penguin contributions limit the theoretical accuracy of the 
extracted value of $\sin\phi_q$ through terms of 
${\cal O}(R_b\lambda^3)={\cal O}(4\times 10^{-3})$ \cite{LHC-Report,FM}. 
Consequently, the simple determinations of this quantity with the help of
(\ref{sin-phi-det}) are theoretically cleaner by one order of magnitude. 
For approaches to control the penguin uncertainties in $B_d\to J/\psi 
K_{\rm S (L)}$ and $B_s\to J/\psi \phi$ through flavour-symmetry arguments, 
see \cite{pen-cont}. These issues may become important in the era of the LHC.

Needless to note, it will be very challenging to achieve such a 
tremendous accuracy in practice. To this end, we must also keep an eye 
on CP violation in $B^0_q$--$\overline{B^0_q}$ mixing, which enters at the 
$10^{-4}$ level in the Standard Model and can be probed through 
``wrong-charge'' lepton asymmetries \cite{RF-PHYS-REP}; these
effects may be enhanced sizeably through new physics. As we have 
noted above, a similar comment applies to $D^0$--$\overline{D^0}$ mixing
and CP violation in $D$ decays. Moreover, also indirect CP violation in the 
neutral kaon system has to be taken into account if $K_{\rm S}$ or 
$K_{\rm L}$ are involved. However, these effects do not lead to serious 
problems and can be included through the corresponding 
experimental data. 

The strategy provided by (\ref{sin-phi-det}) is particularly interesting for 
the $B_s$-meson case. If $\phi_s=\phi_s^{\rm SM}=-2\lambda^2\eta\approx 
-3\times 10^{-2}$, we may determine $\phi_s$ with the help of 
$B_s\to D_\pm K_{\rm S (L)}$ modes with a theoretical accuracy of 
${\cal O}(1\%)$. On the other hand, as emphasized in \cite{LHC-Report}, 
using the time-dependent angular distribution of the $B_s\to J/\psi \phi$ 
decay products, the theoretical accuracy of $\phi_s$ is limited by penguin 
contributions to ${\cal O}(10\%)$. An accurate determination of $\phi_s$ is 
very important, since it is essentially given by the Wolfenstein parameter 
$\eta$ within the Standard Model, measuring the height of the unitarity 
triangle. An alternative approach to determine this parameter is provided 
by the rare kaon decay $K_{\rm L}\to\pi^0\nu\overline{\nu}$ 
\cite{RF-PHYS-REP,Andr-02}, which would offer an interesting consistency check.

Let us finally note that the $\overline{B^0_d}\to D^0\pi^0$ mode has
already been observed by the Belle, CLEO and BaBar collaborations, with
branching ratios at the $3\times 10^{-4}$ level \cite{Bbar-D0pi0}. Rescaling
them by $\lambda^2$, we obtain the estimate $\mbox{BR}(\overline{B^0_d}\to 
D^0 K_{\rm S})={\cal O}(10^{-5})$. Interestingly, the $\overline{B^0_d}\to 
D^0 \overline{K^0}$ channel has very recently been observed for the first
time by the Belle collaboration, with the branching ratio 
$(5.0^{+1.3}_{-1.2}\pm0.6)\times10^{-5}$ \cite{Belle-BdDK-obs}.

\section{Conclusions}\label{sec:concl}
In summary, we have shown that $B_d\to D_\pm K_{\rm S (L)}$ and 
$B_s\to D_\pm \eta^{(')}, D_\pm \phi, ...$\ modes provide a very efficient 
strategy to determine $\gamma$ in a theoretically clean and essentially
unambiguous manner. Here we mean by ``efficient'' that only the ``untagged'' 
asymmetry $\Gamma_{+-}^{f_s}$ and the mixing-induced CP-violating observables 
$\langle S_{f_s}\rangle_+$ and $\langle S_{f_s}\rangle_-$ of these
colour-suppressed decays have to be measured, which satisfy the simple,
{\it exact} relation (\ref{final-result}), allowing us to extract 
$\tan\gamma$. Interestingly, first information about $\gamma$ can already 
be obtained from $\Gamma_{+-}^{f_s}$ with the help of 
$|\cos\gamma|\geq|\Gamma_{+-}^{f_s}|$, which can straightforwardly 
be converted into bounds on $\gamma$. 

It is useful to compare this new strategy with other approaches to extract 
weak phases from decays of the kind $B_d\to D K_{\rm S}$. In \cite{GroLo}, 
it was pointed out that these modes allow a clean determination of 
$\sin\phi_d\stackrel{{\rm SM}}{=}\sin2\beta$ and 
$\sin(\phi_d+2\gamma)\stackrel{{\rm SM}}{=}-\sin2\alpha$, leaving fourfold 
discrete ambiguities for the values of $\beta$ and $\alpha$. To this end, 
time-dependent rate measurements of $B_d\to D^0 K_{\rm S}$ and 
$B_d\to \overline{D^0} K_{\rm S}$ processes are complemented with a complex 
triangle construction involving the $B_d\to D_+ K_{\rm S}$ channel. In 
\cite{KaLo}, it was then shown that this method is affected by subtle 
interference effects between $D^0\to\pi^+K^-$ and $\overline{D^0}\to\pi^+K^-$, 
which are Cabibbo-favoured and doubly Cabibbo-suppressed decay 
processes, respectively. The approach proposed in \cite{KaLo} to extract 
$\beta$ and $\gamma$ from time-dependent $B_d\to D^0 K_{\rm S}$ and 
$B_d\to \overline{D^0} K_{\rm S}$ measurements ($B_d\to D_\pm K_{\rm S}$
is not used here), taking into account the $D$-decay interference effects, 
can no longer be implemented in an analytical manner, and leaves a 16-fold 
discrete ambiguity for the extraction of $\beta$, $\gamma$ and two strong 
phases. The essentially unambiguous extraction of $\gamma$ through decays 
of the kind $B_d\to D_\pm K_{\rm S}$ proposed above is obviously much simpler
and in this sense more efficient. Moreover, it is of advantage not to 
employ the Standard-Model expression $\phi_d=2\beta$, since this phase may 
well be affected by new-physics contributions to $B^0_d$--$\overline{B^0_d}$ 
mixing, and can straightforwardly be determined separately. 

The formalism developed for $B_d\to D_\pm K_{\rm S (L)}$ and 
$B_s\to D_\pm \eta^{(')}, D_\pm \phi, ...$\ modes can also be applied 
to decays of the kind $B_d\to D_\pm\pi^0, D_\pm\rho^0, ...$\ and 
$B_s\to D_\pm K_{\rm S (L)}$, where the interference effects between the 
$\overline{B^0_q}\to \overline{D^0} f_d$ and $\overline{B^0_q}\to D^0 f_d$ 
decay paths are doubly Cabibbo-suppressed, thereby reducing the prospects 
of these modes to determine $\gamma$. However, these channels allow 
interesting determinations of $\sin\phi_q$. In comparison with the 
conventional methods provided by $B_d\to J/\psi K_{\rm S (L)}$ and 
$B_s\to J/\psi \phi$, these extractions are theoretically cleaner by one 
order of magnitude. A more comprehensive analysis of $B_q\to D f_{s,d}$ 
modes is given in \cite{RF-BDf-long}. There we also have a closer look at 
the interesting case where the neutral $D$ mesons are observed through 
decays into CP non-eigenstates. We look forward to detailed experimental 
studies of the corresponding decays.

\end{document}